\documentclass[preprint,12pt]{elsarticle}

\usepackage{amssymb,amsmath,amsfonts,esint,xcolor}
\biboptions{sort&compress}
\usepackage{tabularx}
\usepackage{graphicx}
\usepackage[ruled,linesnumbered]{algorithm2e}

\journal{Journal of Colloid and Interface Science}
\begin{document}
\begin{frontmatter}



\title{Scattering Insights into Shear-Induced Scission of Rod-like Micelles}

\author[NTHU,NCTS]{Guan-Rong Huang}
\affiliation[NTHU]{organization = {Department of Engineering and System Science}, 
addressline={National Tsing Hua University}, 
city={Hsinchu},
postcode={30013}, 
country={Taiwan}}
\affiliation[NCTS]{organization = {Physics Division}, 
addressline={National Center for Theoretical Sciences}, 
city={Taipei},
postcode={10617}, 
country={Taiwan}}

\author[NIST]{Ryan P. Murphy}
\affiliation[NIST]{organization={NIST Center for Neutron Research},
            addressline={National Institute of Standards and Technology}, 
            city={Gaithersburg},
            postcode={20878}, 
            state={Maryland},
            country={United States}}

\author[ILL]{Lionel Porcar}
\affiliation[ILL]{organization={Institut Laue-Langevin}, addressline={ 71 Avenue des Martyrs, B.P. 156, F-38042 Grenoble Cedex 9}, country={France}}

\author[NSD]{Chi-Huan Tung}
\author[NSD]{Changwoo Do}
\author[NSD]{Wei-Ren Chen}
\affiliation[NSD]{organization={Neutron Scattering Division},
            addressline={Oak Ridge National Laboratory}, 
            city={Oak Ridge},
            postcode={37831}, 
            state={Tennessee},
            country={United States}}

\begin{abstract}
\begin{flushleft}
\emph{Hypothesis}\\
\hfill\break

\textcolor{black}{Understanding the scission of rod-like micelles under mechanical forces is crucial for optimizing their stability and behavior in industrial applications. This study investigates how micelle length, flexibility, and external forces interact, offering insights into the design of micellar systems in processes influenced by mechanical stress. Although significant, direct experimental observations of flow-induced micellar scission using scattering techniques remain scarce.}\\

\hfill\break
\emph{Experiments and Simulations}\\
\hfill\break

\textcolor{black}{Small angle neutron scattering (SANS) is used to explore the shear response of aqueous cetyltrimethylammonium bromide (CTAB) solutions with sodium nitrate. Rheological tests show shear thinning with no shear banding, ensuring a uniform flow field for reliable interpretation of scattering data. As shear rate increases, the scattering spectra show angular distortion, which is analyzed using spherical harmonic decomposition to characterize flow-induced scission and micelle orientation under shear.}\\

\hfill\break
\emph{Findings}\\
\hfill\break

\textcolor{black}{Two analysis steps are used: a model-independent spectral eigendecomposition reveals a decrease in micellar length, while regression analysis quantifies the evolution of the length distribution and mean length with shear rate. Additionally, micelle alignment increases with shear, quantified by the orientational distribution function. These findings provide experimental evidence for flow-induced alignment and scission, offering a new framework for understanding shear-induced phenomena in micellar systems.}

\end{flushleft}
\end{abstract}
\clearpage
\begin{graphicalabstract}
\includegraphics[width=\linewidth]{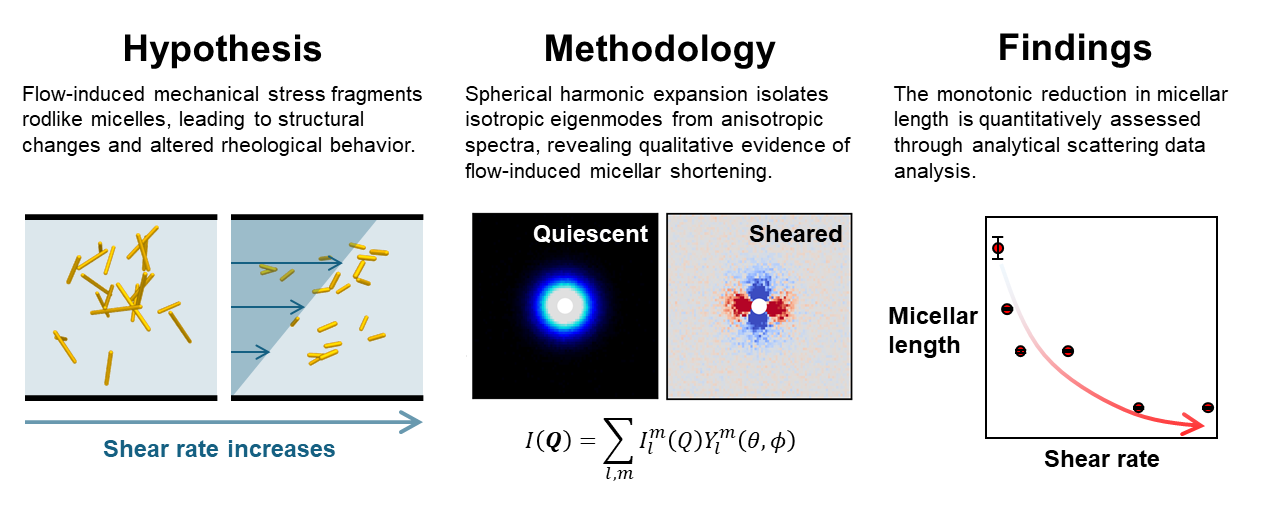}
\end{graphicalabstract}



\end{frontmatter}

\clearpage
\section{Introduction}
\label{sec:1}

\textcolor{black}{The structural and dynamic behavior of elongated, rod-like micelles under shear flow is fundamental to understanding the fluid dynamics and material properties of colloidal and self-assembled systems, which play a pivotal role in fields such as nanomedicine, catalysis, and environmental science. These systems exhibit unique, flow-sensitive properties, making the study of their deformation and transformations under shear crucial for advancing colloid and interface science. This understanding has far-reaching applications in industries such as food processing, polymer manufacturing, and drug delivery.}

\textcolor{black}{One significant flow-induced phenomenon in cylindrical micellar systems is the reduction in contour length, which directly affects the distribution and alignment of micelles. This, in turn, leads to changes in viscosity, elasticity, and shear-thinning behavior, profoundly influencing the macroscopic rheological properties of micellar solutions.} 

\textcolor{black}{Computational and theoretical studies \cite{Cates1, Cates2, Warr1, Walker, Lequeux1, Makhloufi, Fisher, Lequeux2, Rothstein1, Vasquez, Rothstein2, Padding, Sureshkumar, Larson, Rothstein3, Koide1, Koide2} have provided valuable insights into the molecular mechanisms underlying micellar scission and its structural consequences. These contributions are critical for predicting micellar behavior under flow and for designing systems with tailored rheological properties in industrial applications. Alongside these advances, small-angle neutron scattering (SANS) has emerged as an indispensable experimental technique for probing the structural properties of elongated micelles. SANS complements molecular simulations by providing detailed experimental data on the conformational properties of cylindrical micelles, particularly in equilibrium systems \cite{ILL_16}.}

\textcolor{black}{Building on the pioneering work of Hayter and colleagues \cite{Hayter3, Hayter1, Hayter2}, extensive efforts have been made to investigate the non-equilibrium conformations of elongated micelles under mechanical forces, primarily through rheo-SANS experiments \cite{May, Kalus1, Kalus2, Kalus3, Berret, Foerster, Wagner1, Wagner2, Rothstein2, Wagner3, Lettinga, Wagner4, Weigandt, Yun, Tabor1, Lionel2, Tabor2, Helgeson}. These studies have significantly advanced our understanding of micellar alignment and its effects on rheological properties. However, the direct observation and quantitative analysis of micellar scission—a key process for understanding flow behavior—remains elusive, despite decades of research.} 

\textcolor{black}{The primary challenge lies in disentangling the scattering signature of micellar scission from the broader spectral anisotropy caused by factors such as orientational and length distributions. These complexities arise from the interplay between thermodynamic forces and fluid mechanical interactions that drive micellar systems out of equilibrium. Addressing this challenge remains a central research objective, as highlighted by recent studies \cite{Helgeson}.}

\textcolor{black}{In this work, we present a robust spectral analysis framework based on spherical harmonic decomposition to study the small-angle scattering intensity profiles of rod-like cetyltrimethylammonium bromide (CTAB) micelles under shear. Using the mathematical principles of orthogonal basis expansion, the collected spectra are decomposed into spherical harmonic components, which are eigenfunctions of angular momentum operators and well-suited for analyzing orientational distortions \cite{Schiff, huang2025}. This approach effectively isolates the isotropic scattering intensity component, \(I_0^0(Q)\), enabling the direct quantification of micellar length variations and their evolution under flow. Spherical harmonic components with low angular symmetry provide insights into the micellar orientation distribution function, revealing hydrodynamic effects on these rod-like micelles.}

\textcolor{black}{Our results demostrate that increased shear rates induce a distinct spectral anisotropy and a reduction in \(I_0^0(Q)\) within the flow-velocity gradient plane, with angular asymmetry emerging at shear rates above 500 s\(^{-1}\). The decrease in the isotropic component, along with the observed mirror asymmetry, suggests scission events that alter the colloidal stability and polydispersity of micelles in response to the applied flow fields. This reduction in the isotropic component reflects a decrease in effective micellar length, while the asymmetry indicates the alignment of micelles. Using the Schulz distribution model, we capture the statistical distribution of micellar lengths, showing a decrease from an average of 400 \AA\ at rest to 150 \AA\ under high shear of 3000 s\(^{-1}\). Additionally, the alignment of rod-like micelles is progressively enhanced with increasing shear rates.}

\textcolor{black}{This quantitative framework provides a robust methodology for studying shear-induced transformations in soft matter systems, offering new insights into the relationship between flow dynamics and micellar structure. Beyond advancing our understanding of micellar behavior, this work establishes a foundation for exploring structural evolution in colloidal systems under deformation. These insights have broad implications for materials science, catalysis, and other areas of applied colloid science.}

\section{Materials and Rheology}
\label{sec:2}

Solutions were prepared by weighing CTAB (cetyltrimethylammonium bromide, as-recieved, 99\% purity, MilliporeSigma) and sodium nitrate (NaNO$_3$, as-received, 99\% purity, MilliporeSigma) in pure water at specified molar concentrations with a constant molar ratio of NaNO$_3$:CTAB of 3:1. For simplicity, the concentration identified in the text and figures refer to the surfactant molar concentration (e.g. 30 mM refers to 30 mM CTAB with 90 mM NaNO$_3$). Samples were prepared in ultra-purified water (H$_2$O, 18.2 M$\Omega$-cm resistivity) for rheological characterization, and were prepared in deuterated water (D$_2$O, 99.9\% purity, Cambridge Isotope Laboratories, Inc.) to provide suitable coherent contrast for small angle neutron scattering measurements. Samples were mixed in a heated incubation shaker at 40~$^{\circ}$C for several hours until the solution was visibly clear and homogeneous. No visible precipitation was observed when operating at or above 25~$^{\circ}$C, but delayed crystallization was observed if the solution was kept at 20~$^{\circ}$C or less (meta-stable region). Thus, the rheology and scattering measurements were performed in the apparent thermodynamic stable region at 25~$^{\circ}$C.

\begin{figure}[h!]
\centerline{
  \includegraphics[scale=0.25]{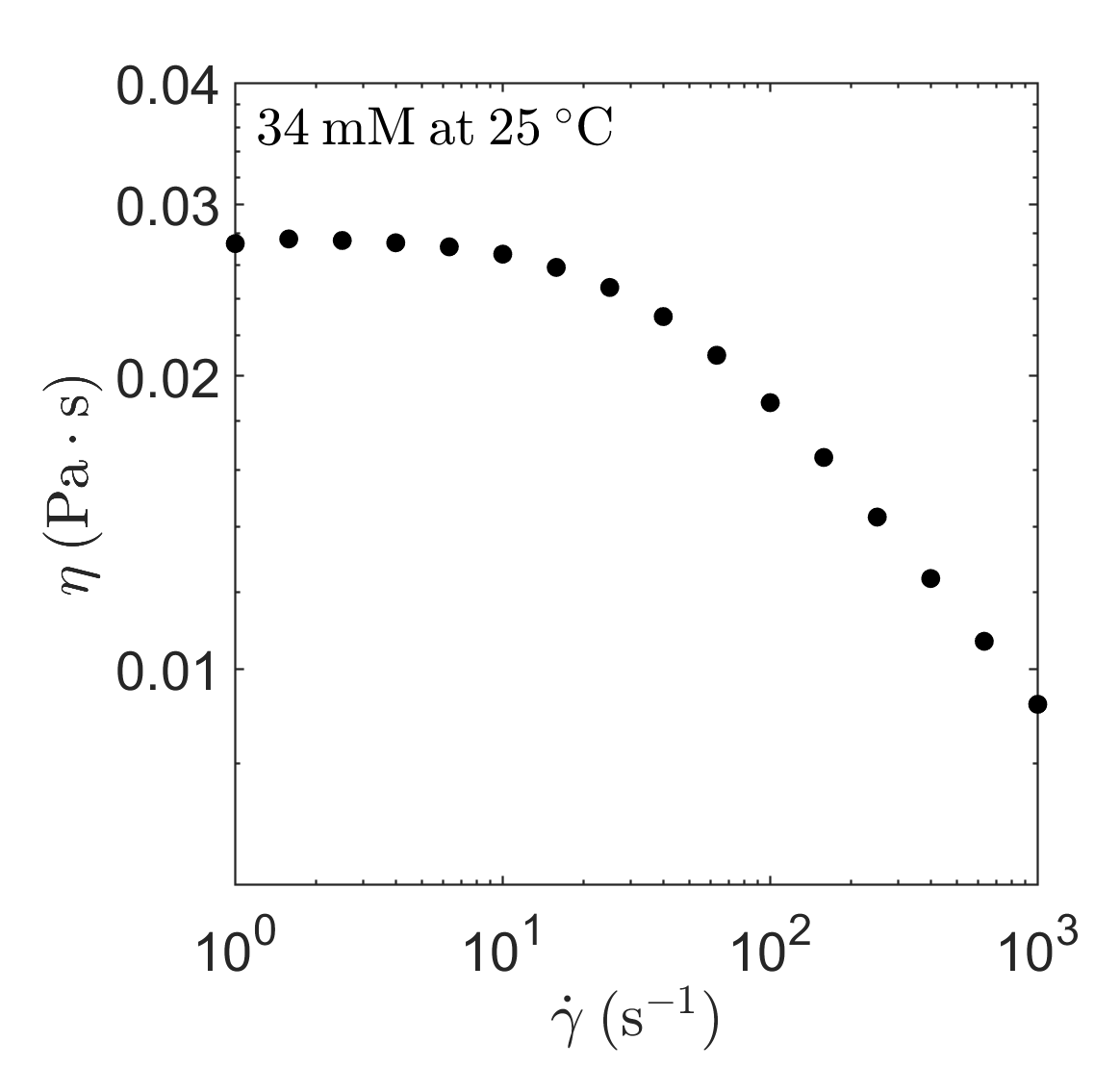}
}  
\caption{Steady-shear viscosity measurements as a function of \textcolor{black}{shear rate} ($\dot{\gamma}$) over a range of 1 s$^{-1}$ to 1000 s$^{-1}$. Each data point represents an average over a 3-minute interval. In the concentration and temperature ranges of interest for SANS measurements, no significant time-dependent fluctuations or oscillations in the measured stress were observed, indicating the absence of shear banding, vorticity banding, or elastic instabilities under the applied flow field.   
}  
\label{fig:1}
\end{figure}

Rheological measurements were performed using a strain-controlled ARES-G2 rheometer (Waters-TA Instruments) equipped with a rotating concentric cylinder sample geometry (30 mm ID aluminium cup, 29 mm OD stainless steel bob), advanced Peltier system for accurate temperature control to within 0.1~$^{\circ}$C of the setpoint, and a water-filled solvent trap to minimize water evaporation throughout repeated measurements. The steady-shear viscosity was measured both in the increasing and decreasing shear rate ($\dot{\gamma}$), spanning from approximately 1 s$^{-1}$ to 1000 s$^{-1}$ with an approximate 3 minute averaging period at each point. \textcolor{black}{In the concentration and temperature ranges of interest for SANS measurements, the measured stress did not show any large time-dependent fluctuations or oscillations, which would otherwise be indicated of shear banding, vorticity banding, or elastic instabilities due to the imposed flow field.}

\textcolor{black}{The relaxation time is defined as $\Lambda = 2 \pi \omega_c ^{-1}$, where $\omega_c$ (rad/s) is the interpolated crossover frequency in the linear viscoelastic regime. In the examined case, $\Lambda =~0.10~\pm~0.04$~s at the measured surfactant concentration (34 mM CTAB) and temperature (25~$^{\circ}$C). This value is similar to previous studies \cite{Helgeson2010}. The dimensionless Weissenberg number (Wi) is defined as $Wi = \Lambda \dot{\gamma}$, which is later plotted on the x-axis in place of the shear rate to enable better comparison with other rod-like or worm-like micelle systems that have different longest relaxation times.}

\section{Small Angle Neutron Scattering Experiment}

\textcolor{black}{The microscopic structure of the sheared micellar solution was investigated using flow-SANS measurements with the 1-2 shear cell Couette flow geometry \cite{Gurnon}.} The experimental design was structured around three principal directions aligned with the applied shear flow: the flow direction (\(\mathbf{v}\), referred to as the 1-direction), the velocity gradient direction (\(\nabla \mathbf{v}\), referred to as the 2-direction), and the vorticity direction (\(\mathbf{x} = \nabla \times \mathbf{v}\), referred to as the 3-direction). SANS measurements were conducted in the flow–velocity gradient plane (\(\mathbf{v}\)–\(\nabla \mathbf{v}\), or the 1–2 plane), as this plane provides the most critical insights into the relationship between flow-induced structural changes and shear viscosity \cite{Soloman}. \textcolor{black}{The CTAB concentration and temperature were maintained at constant values of 34 mM and 25~$^{\circ}$C, respectively, throughout all SANS measurements. The system was selected to be in the dilute regime, where inter-micellar spatial correlations are negligible.} 
\textcolor{black}{We also have ensured that at the highest shear rate probed within the 1-2 shear cell, the flow behavior remains laminar as both dimensionless Reynolds number (Re) and Taylor number (Ta) defined for Taylor-Couette flow geometry with a rotating inner cylinder as $Re = v_i \rho d \eta^{-1}\approx500$, and $Ta = d^{1/2} R_i^{-1/2} Re \approx70$ remain below the critical onset values known for inertia-driven turbulence ($Re<2000$ and $Ta<200$). Here, $v_i$ is the characteristic angular velocity of the moving inner wall ($v_i = 3$~m~s$^{-1}$ at the maximum examined shear rate), $R_i$ is the exterior radius of the moving inner wall ($R_i = 0.051$~m), $d$ is the gap between the inner and outer wall where the sample resides ($d = 0.001$~m), $\rho$ is the density of the sample ($\rho=1100$~kg~m$^{-3}$), and $\eta$ is the sample viscosity ($\eta = 0.0065$~Pa~s at the maximum examined shear rate).}

By focusing on the 1-2 plane, the experiment facilitates a more detailed analysis of how shear influences the microstructure, which in turn affects the rheological properties of the system.

\begin{figure}[h!]
\centerline{
  \includegraphics[width=\linewidth]{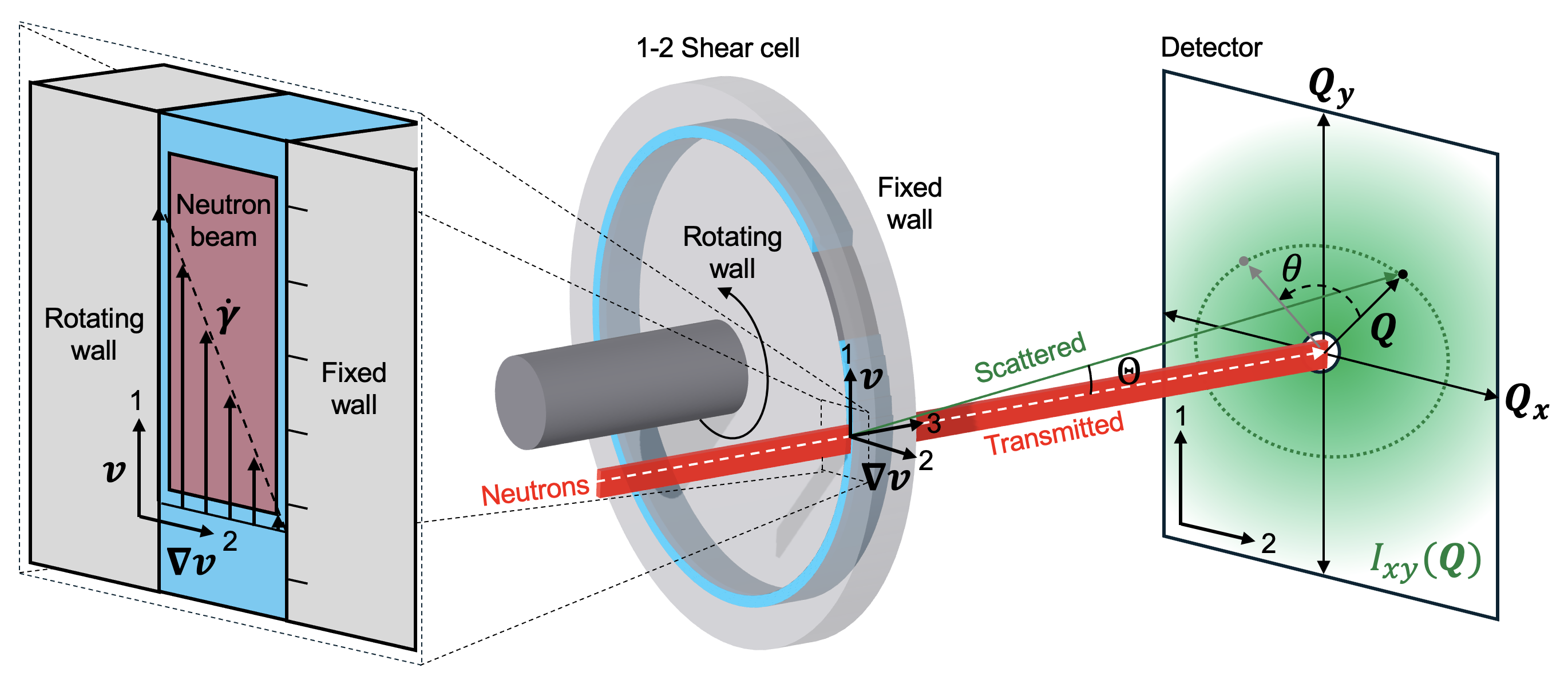}
}  
\caption{\textcolor{black}{Schematic of the SANS experiment conducted under Couette geometry. The directions of flow (\(v\)), velocity gradient (\(\nabla v\)), and vorticity (\(\nabla \times v\)) are indicated as 1, 2, and 3, respectively. Two-dimensional SANS spectra were acquired in the flow–velocity gradient plane (\(1{-}2\) or \(\nabla v{-}v\)).}
 }  
\label{fig:2}
\end{figure}

The SANS experiments were conducted using the D22 diffractometer at the Institut Laue-Langevin (ILL). The instrument provides an extensive dynamic range, making it well-suited for these measurements. Data acquisition was performed with a collimation distance of 17.6 m and a source aperture of 40 mm in diameter, utilizing a neutron wavelength of \(6 \, \text{\AA}\). No beam stop was employed due to the low beam intensity, which resulted from the use of a \(3 \, \text{mm} \times 0.75 \, \text{mm}\) sample slit aperture to probe the flow-gradient velocity plane of the shear cell. \textcolor{black}{ A schematic illustration of the experimental setup is presented in Fig.~\ref{fig:2}.}

The scattering vector magnitude is defined as \(Q=4\pi\lambda^{-1}\sin(\Theta/2)\), where \(\lambda\) is the neutron wavelength and \(\Theta\) is the scattering angle. To cover the full range of scattering vector magnitudes, two measurements were performed while maintaining a constant collimation distance. The first measurement used the back detector positioned at 17.6 m, covering a \(Q\) range from \(0.0025 \, \text{\AA}^{-1}\) to \(0.04 \, \text{\AA}^{-1}\). The second measurement utilized the side detector placed at 1.4 m, centered around the neutron beam, covering a \(Q\) range from \(0.02 \, \text{\AA}^{-1}\) to \(0.43 \, \text{\AA}^{-1}\). Data reduction was performed with Grasp following the standard procedure and using the direct beam to provide the absolute scale \cite{Grasp}.

\section{Results and Discussion}
\label{sec:3}

\subsection{\textcolor{black}{Basis Decomposition of Scattering Spectra Through Real Spherical Harmonic Expansion}}

The coherent scattering intensity distribution of mechanically driven CTAB micellar systems, which exhibit cylindrical symmetry, is known to display angular asymmetry under flow. The anisotropic scattering intensity, \( I(\mathbf{Q}) \), can be expressed as a linear expansion in terms of real spherical harmonics (RSH):
\begin{equation}
    I(\mathbf{Q}) = \sum_{l,m} I_l^m(Q) Y_l^m(\theta, \phi),
\label{eq:3.1}
\end{equation}
where \( Y_l^m(\theta, \phi) \) denotes the RSH of degree \( l \) and order \( m \), with \( \theta \) and \( \phi \) representing the polar and azimuthal angles in reciprocal space, respectively. In general, \( I(\mathbf{Q}) \) is a three-dimensional function expressed in spherical coordinates, complicating the direct reconstruction of the full intensity distribution from a two-dimensional scattering spectrum.

However, by exploiting the symmetry imposed by shear flow and defining \( \theta \) as the angle between \( \mathbf{Q} \) and the flow direction (the \( x \)-axis), this complexity can be reduced. A suitable rotational transformation restores axial symmetry to \( I(\mathbf{Q}) \) . Under these conditions, the scattering intensity depends solely on \( Q \) and the polar angle \( \theta \), eliminating the dependence on the azimuthal angle \( \phi \). Consequently, only terms with \( m = 0 \) contribute in Eqn.~(\ref{eq:3.1}), reducing the RSH expansion to a series of Legendre polynomials. This simplification enables the reconstruction of the three-dimensional anisotropic intensity using a single two-dimensional scattering spectrum.

In this study, the RSH expansion is carried out on the flow-velocity gradient plane, with a specific tilt to recover the axial symmetry of the scattering intensity \cite{Huang3}. This symmetry is inherently linked to the molecular shape and the flow field. For spectral analysis, the following formula is employed:
\begin{equation}
    I_l^0(Q) = \frac{1}{2} \int_0^\pi \mathrm{d}\theta \sin{(\theta + \theta_t)} Y_l^0(\theta + \theta_t) I_{xy}(\mathbf{Q}),
\end{equation}
where \( \theta_t \) is the tilt angle relative to \( Q_x \) and \( I_{xy}(\mathbf{Q}) \) is the scattering spectra on the 1-2 plane in reciprocal space. For the isotropic component of \( I(\mathbf{Q}) \), the equation simplifies to:
\begin{equation}
    I_0^0(Q) = \frac{1}{2} \int_0^\pi \mathrm{d}\theta \sin{(\theta + \theta_t)} I_{xy}(\mathbf{Q}).
\end{equation}

These equations highlight the significance of measurements on the 1-2 plane for spectral analysis. First, axial symmetry can be restored on the 1-2 plane by rotating it along the \( Q_z \)-axis at the tilt angle \( \theta_t \), which can be directly determined on the 1-2 plane. As previously reported \cite{Huang3, Lionel1}, the tilt angle \( \theta_t \) is not discernible from scattering spectra on the 1-3 and 2-3 planes. Second, the polar angle on the 1-2 plane spans from \( 0 \) to \( \pi \), ensuring the orthogonality of \( Y_l^0(\theta + \theta_t) \). On the 2-3 plane, the polar angle is \( \pi/2 \), rendering \( Y_l^0 \) independent of \( \theta \), while on the 1-3 plane, \( Y_l^0 \) is a function of \( \cos{\theta_t} \cos{\theta} \). Thus, the orthogonality of \( Y_l^0 \) is lost on the 1-3 and 2-3 planes. This orthogonality decouples each RSH component of \( I(\mathbf{Q}) \), simplifying subsequently the analysis. As a result, structural parameters of rod-like micellar molecules can be extracted from model fitting with the decoupled RSH components on the 1-2 plane.

\subsection{\textcolor{black}{Experimental Validation of Micellar Scission Using Spectral Eigendecomposition}}

Among all RSH components of \( I(\mathbf{Q}) \), the isotropic component, \( I_0^0(Q) \), is the most relevant for analyzing the structural parameters of rod-like molecules for several reasons. First, once the length and cross-sectional radius of the rods are given, \( I_0^0(Q) \) remains invariant under rotational transformation, regardless of the rod orientation distribution function. In the case of rigid rods, both the rod length and radius are unaffected by externally applied mechanical fields, leading to the invariance of \( I_0^0(Q) \) and the radius of gyration. Consequently, changes in rod length can be directly determined through variation in \( I_0^0(Q) \). Furthermore, the primary contribution to the overall scattering intensity comes from \( I_0^0(Q) \), and the ratio of its magnitude to its associated error is minimal, typically within 5\%. Therefore, focusing on \( I_0^0(Q) \) provides statistically robust results in scattering data analysis.

\begin{figure}[h!]
\centerline{
  \includegraphics[width=\linewidth]{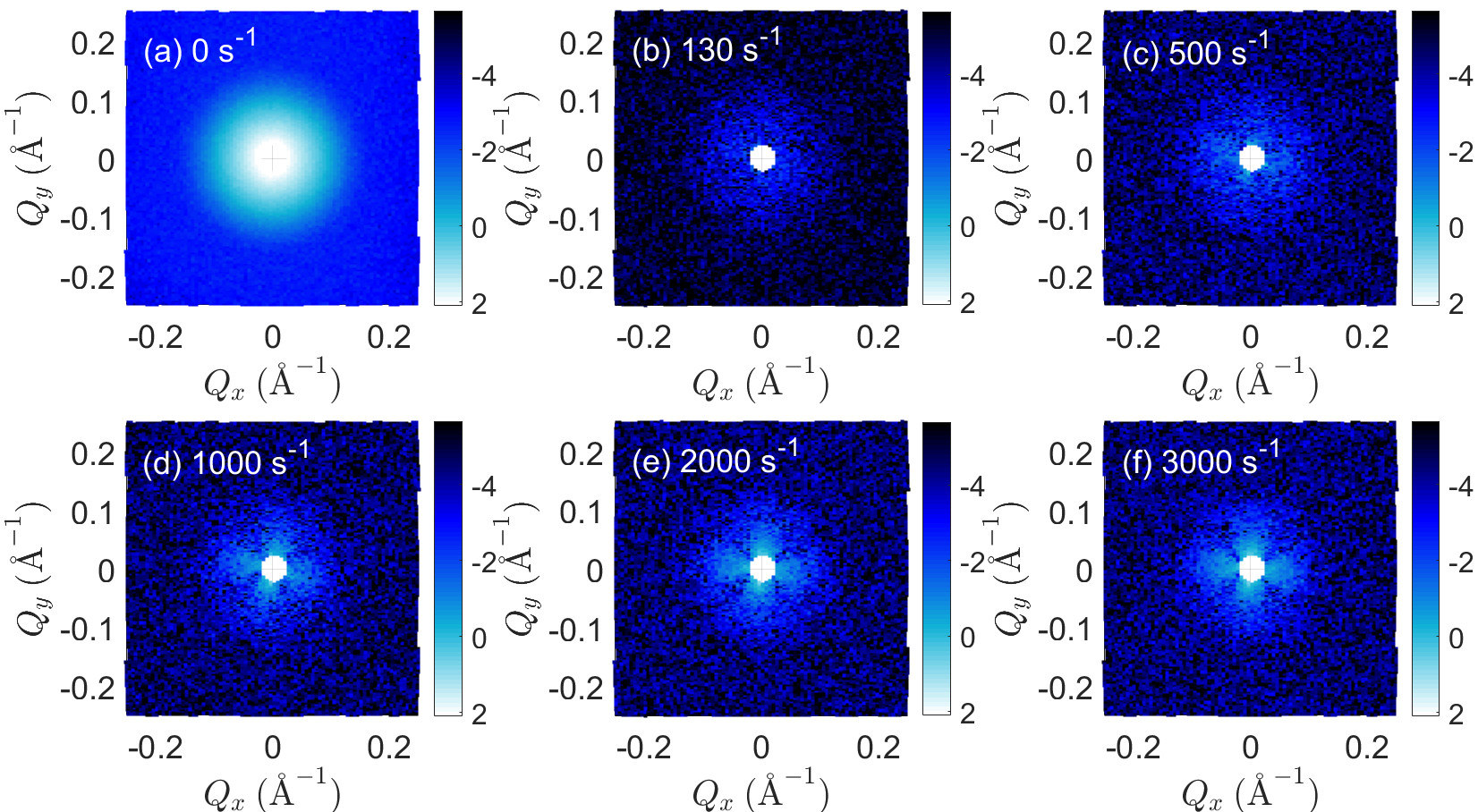}
}  
\caption{Panel (a) shows the 2D scattering intensity, $I_{xy}(\mathbf{Q})$, of the aqueous \textcolor{black}{CTAB} solution in its quiescent state. Panels (b-f) depict the difference in $I_{xy}(\mathbf{Q})$ for the \textcolor{black}{CTAB} solution under steady shear at shear rates $\dot{\gamma} = 130$, 500, 1000, 2000, and 3000 s$^{-1}$, relative to the quiescent state presented in Panel (a). The circle at the center of each spectrum marks the position of the beam stop. \textcolor{black}{It is important to note that the apparent rotation of the two-dimensional spectra reflects the averaged orientation of micelles, which aligns progressively more closely with the flow direction as shear rates increase \cite{Huang3}. These measurements were taken using the same experimental setup, although at varying shear rates.}
}  
\label{fig:3}
\end{figure}

Figure~\ref{fig:3} shows the projection of $I(\mathbf{Q})$ for rod-like micelles on the flow-velocity gradient plane under Couette flow. In this setup, the flow direction is along the $y$-axis, with the velocity gradient along the $x$-axis, leading to the quantity being denoted as $I_{xy}(\mathbf{Q})$. 

\textcolor{black}{Figure~\ref{fig:3}(a) illustrates the intensity distribution in the quiescent state, where no flow fields are imposed, highlighting the expected angular isotropy along the azimuthal angle.} Figs.~\ref{fig:3}(b-f) depict the evolution of $I_{xy}(\mathbf{Q})$, relative to the quiescent state, as a function of the shear rate, $\dot{\gamma}$. Notably, a marked change in the intensity distribution, exhibiting asymmetry about both the vertical and horizontal axes, becomes apparent when $\dot{\gamma} > 500\,\text{s}^{-1}$. This feature grows more pronounced as $\dot{\gamma}$ increases to $3000\,\text{s}^{-1}$. The emergence of such asymmetry in the scattering pattern indicates that spherical harmonic basis functions with $m=0$, specifically the Legendre polynomials, suffice for an accurate spectral analysis, consistent with the reflection symmetry observed in the intensity distribution.

\begin{figure}[h!]
\centerline{
  \includegraphics[scale = 0.225]{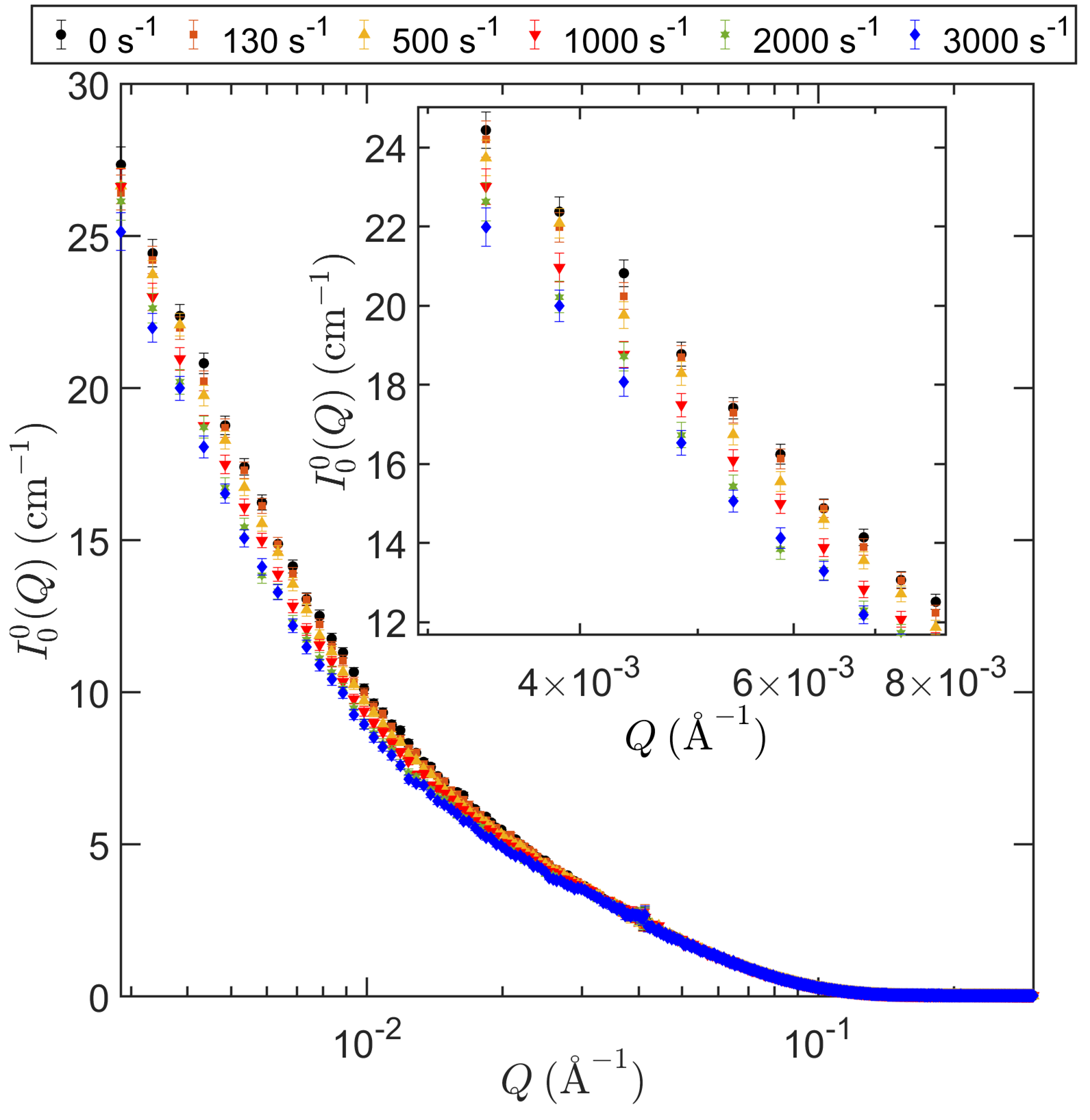}
}  
\caption{$I_0^0(Q)$, extracted from the experimentally measured $I_{xy}(\mathbf{Q})$ through spherical harmonic decomposition, at varying shear rates $\dot{\gamma}$. As $\dot{\gamma}$ increases from 0 to $3000\,\text{s}^{-1}$, a steady decrease in $I_0^0(Q)$ is observed.  
}  
\label{fig:4}
\end{figure}

Figure~\ref{fig:4} displays the isotropic scattering intensity, $I_0^0(Q)$, extracted from the experimentally measured $I_{xy}(\mathbf{Q})$ using the spherical harmonic decomposition, under varying shear rates $\dot{\gamma}$. As $\dot{\gamma}$ increases from 0 to $3000\,\text{s}^{-1}$, a steady decline in the magnitude of $I_0^0(Q)$ is observed in the low-$Q$ region, specifically for $Q < 0.02$~\r{A}$^{-1}$. The inset highlights that the fluctuations in $I_0^0(Q)$ exceed the statistical uncertainty and is gradual with shear rates.  

To rigorously interpret the physical implications of this observation, it is crucial to begin by examining the mathematical framework underlying $I_0^0(Q)$. As shown in \cite{Huang1}, $I_0^0(Q)$ is defined as:
\begin{equation}
I_0^0(Q) = \frac{1}{2\pi^2} \int dr\, r^2 j_0(Qr)g_0^0(r),
\label{eq:4.1}
\end{equation}
where this expression specifically captures the isotropic component of intra-particle spatial correlations \( g_0^0(r) \) in real space and \( j_0(Qr)\) is the spherical Bessel function of the first kind of order \( 0 \). In the case of a rod-like object with cylindrical symmetry, density fluctuations in both real and reciprocal spaces must exhibit equivalent parity. Thus, the most direct approach to understanding these fluctuations is by analyzing the impact of rotational dynamics on the pair correlation function $g_0^0(r)$ in real space, from which reciprocal space behavior follows naturally.

For a freely rotating rigid rod of length $L$, the probability of detecting density fluctuations at a radial distance $r$ from its center-of-mass, denoted by $p(r)$, is directly proportional to $g_0^0(r)$. This relationship can be expressed mathematically as:
\begin{equation}
p(r) \propto dr\, 4\pi r^2 g_0^0(r),
\label{eq:4.2}
\end{equation}
In the regime where $r$ is roughly $< L/2$, and as long as the rod length remains fixed, $p(r)$ is expected to remain significant, consistent with the theoretical considerations presented in \cite{Huang2}. However, any reduction in the rod length results in a discontinuous change in $p(r)$. For example, if the rod length decreases from $L$ to $L/2$, $p(r)$ vanishes quickly in the length scales of $r$ being approximately  $> L/4$.

From this analysis, two critical conclusions emerge, as illustrated by the data in Fig.~\ref{fig:4}. First, the reduction in the magnitude of $I_0^0(Q)$ is a direct indicator of a decrease in rod length. Second, the gradual decay of $I_0^0(Q)$ across the measured $Q$ range and the lack of sharp features strongly suggests that the system possesses a significant degree of polydispersity in rod lengths. Notably, the results shown in Fig.~\ref{fig:4} were obtained using a model-free spectral eigendecomposition approach, further supporting the robustness of these findings.

\begin{figure}[h!]
\centerline{
  \includegraphics[width=\linewidth]{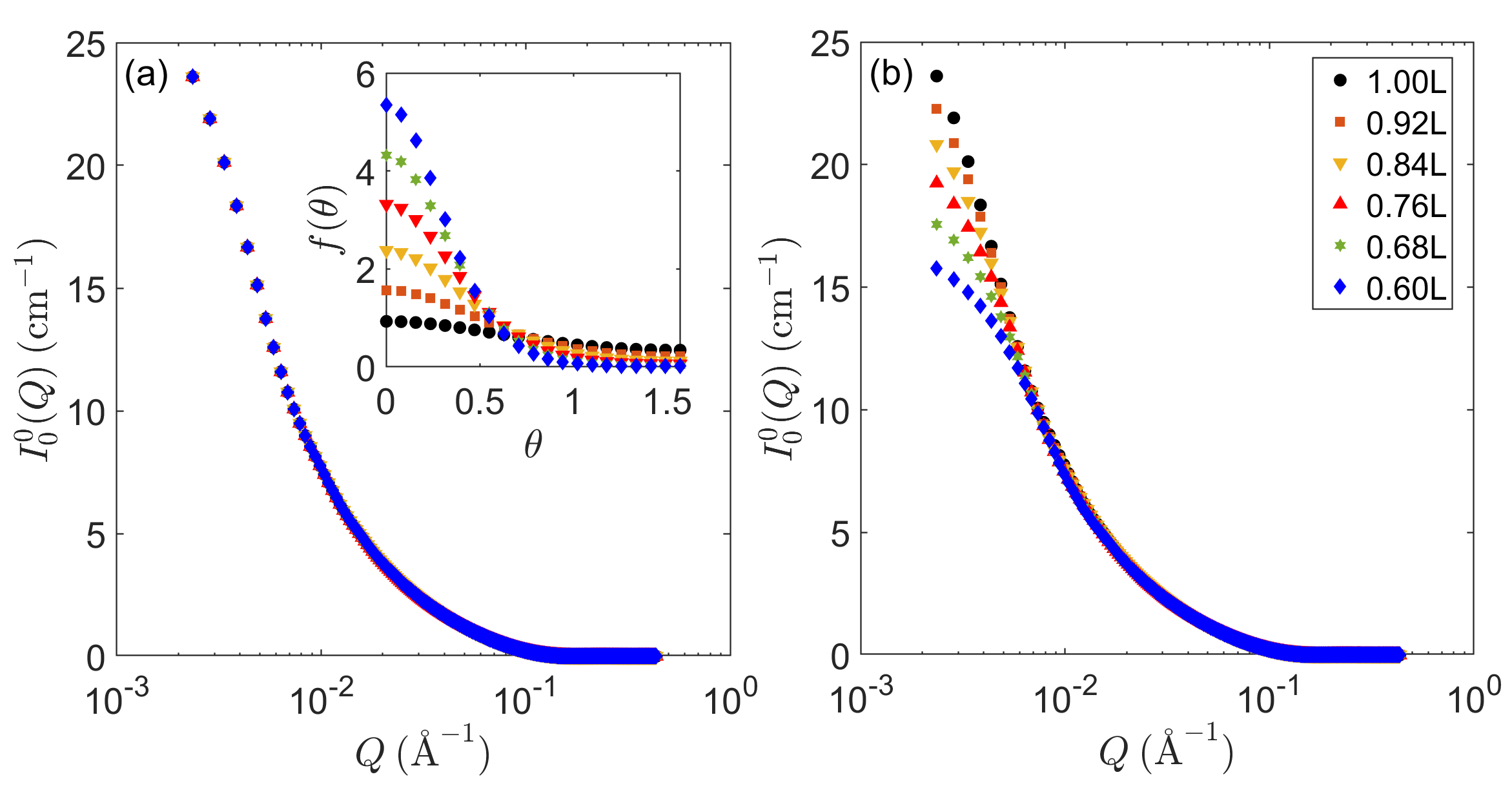}
}  
\caption{Computational benchmarks of a freely rotating rod model with varying orientational distribution functions $f(\theta)$ (shown in the inset), where $\theta$ represents the polar angle relative to the flow direction. (a) For a monodisperse rod length $L = 1050\:\rm{\AA}$ and cross-section radius $R = 21.5\:\rm{\AA}$, the scattering intensity $I_0^0(Q)$ remains invariant across the measured $Q$ range, irrespective of the distribution width of $f(\theta)$. (b) A pronounced decrease in $I_0^0(Q)$ is observed for $Q < 0.01$~\r{A}$^{-1}$ as the rod length is gradually reduced from $L$ to $0.6L$, demonstrating the sensitivity of the scattering intensity to changes in rod length. These results underscore the critical role of rod length in determining the system’s scattering behavior.
}  
\label{fig:5}
\end{figure}

To validate our interpretation of the experimental observations, we performed computational benchmarks using a freely rotating rod model with varying orientational distribution functions $f(\theta)$, where $\theta$ is the polar angle relative to the flow direction. As shown in Fig.~\ref{fig:5}(a), when the rod length $L$ and cross-section radius are constant, the scattering intensity $I_0^0(Q)$ remains unaffected by changes in the distribution width of $f(\theta)$ using the following formula:
\begin{equation}
    I_0^0(Q) = A \int_0^\pi \mathrm{d}\theta \, \sin{\theta} \, \int \mathrm{d}\Omega \, P(\mathbf{Q}, \Omega) f(\Omega),
\end{equation}
where $\Omega$ represents a specific orientation of the rod in real space, $f(\Omega)$ is the orientation distribution function, $P(\mathbf{Q}, \Omega)$ is the scattering function of the rod at orientation $\Omega$, and $A$ is a constant. However, Fig.~\ref{fig:5}(b) demonstrates a systematic decrease in $I_0^0(Q)$ within the $Q < 0.01$~\r{A}$^{-1}$ range as the rod length is monotonically reduced from $L$ to $0.6L$ while keeping the cross-sectional radius \( R \) as a constant.

\subsection{\textcolor{black}{Quantitative Analysis of Micellar Length Distribution}}

\begin{figure}[h!]
\centerline{
  \includegraphics[scale = 0.225]{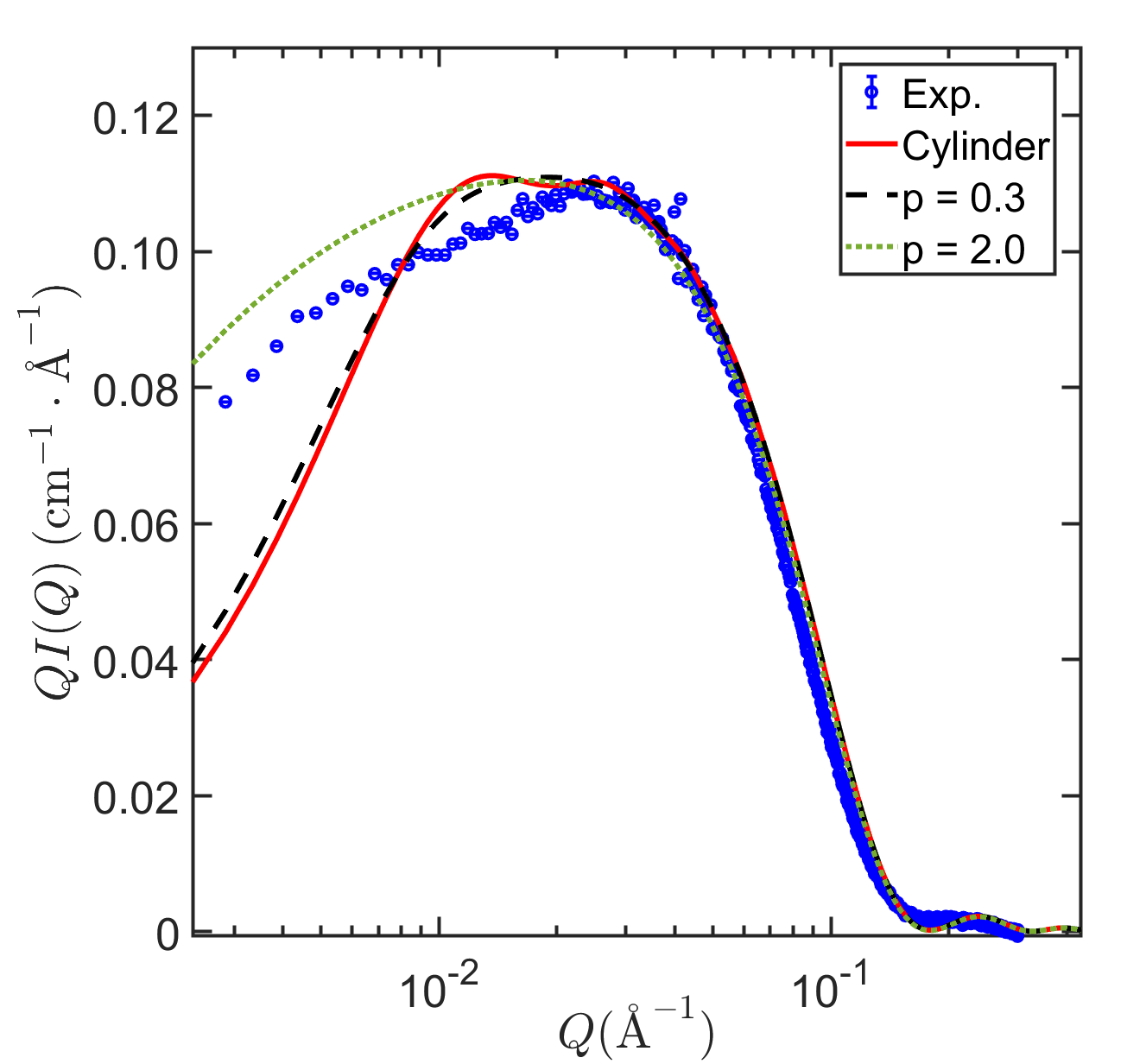}
}
\caption{
Bending rod plot of \(I_0^0(Q)\) for the micellar system in its quiescent state (blue symbols) and for a rigid rod of the length $400\: \text{Å}$ and cross-section radius $21.5\: \text{Å}$ (red curve). In the range \(0.01\, \text{Å}^{-1} < Q < 0.1\, \text{Å}^{-1}\), the product \(QI(Q)\) exhibits a prominent peak. While the overall features of the experimental data align with the monodisperse rod model, a characteristic dip near $Q = 0.02\: \text{Å}^{-1}$, seen in the theoretical benchmark, is missing in the experimental data. This absence is attributed to a smearing effect from the coherent scattering contribution of rods with varying lengths, indicating significant length polydispersity in the system as evidenced by the two corresponding cases of polydispersity index $p = 0.3$ and $2.0$.}
\label{fig:6}
\end{figure}

Fig.~\ref{fig:6} presents the bending rod plot \cite{Magid} of \(I_0^0(Q)\) for the micellar system in its quiescent state (blue symbols). Within the range \(0.01\, \text{Å}^{-1} < Q < 0.1\, \text{Å}^{-1}\), the product \(QI(Q)\) exhibits a prominent peak. For comparison, the bending rod plot for a rigid rod of length \( 400\,\rm{\AA}\), determined through regression analysis, is shown by the red curve. While the overall qualitative features of the experimental data align with the monodisperse rod model, a characteristic dip near the peak at \(Q = 0.02\, \text{Å}^{-1}\) in the theoretical benchmark (red curve) is absent in the experimental data. This discrepancy is attributed to the smearing effect caused by the coherent scattering contribution from rods with different lengths. This observation, along with the difference in the evolution of the magnitude of \(QI(Q)\) in the low-\(Q\) region for systems with polydispersity indices \(p = 0.3\) and \(p = 2.0\), suggests significant length polydispersity in the studied system.

Thus, the results from Figs.~\ref{fig:5} and \ref{fig:6} confirm that the system predominantly consists of rod-like micelles, with $I_0^0(Q)$ being highly sensitive to changes in rod length and polydispersity playing a crucial role in the scattering response.

Although the scattering results in Fig.~\ref{fig:4} strongly indicate flow-induced scission in polydispersed rod-like micelles, a precise quantitative characterization of conformational changes requires additional spectral model fitting. With the micelles modeled as polydispersed rods, we can now perform regression analysis on \(I_0^0(Q)\). This is defined as:

\begin{equation}
I_0^0(Q) = A \int dL \, f(L) P(Q, L, R)  + I_{\text{inc}},
\label{eq:4.3}
\end{equation}
where \(f(L)\) represents the length distribution function, \( A \) is a $Q$-independent constant related to the scattering length density contrast between the micelles and the solvent and the micellar number density, and \(I_{\text{inc}}\) is the incoherent background. \textcolor{black}{The scattering model for polydisperse rods given in Eqn.~\eqref{eq:4.3} describes how rod-like particles of varying sizes scatter neutrons. Mathematically, it averages the scattering contributions over a distribution of rod lengths and orientations, capturing the effects of polydispersity on the overall intensity. Physically, it links the scattering profile to structural features like size, shape, and spatial arrangement of the rods.}

\begin{figure}[h!]
\centerline{
  \includegraphics[width=\linewidth]{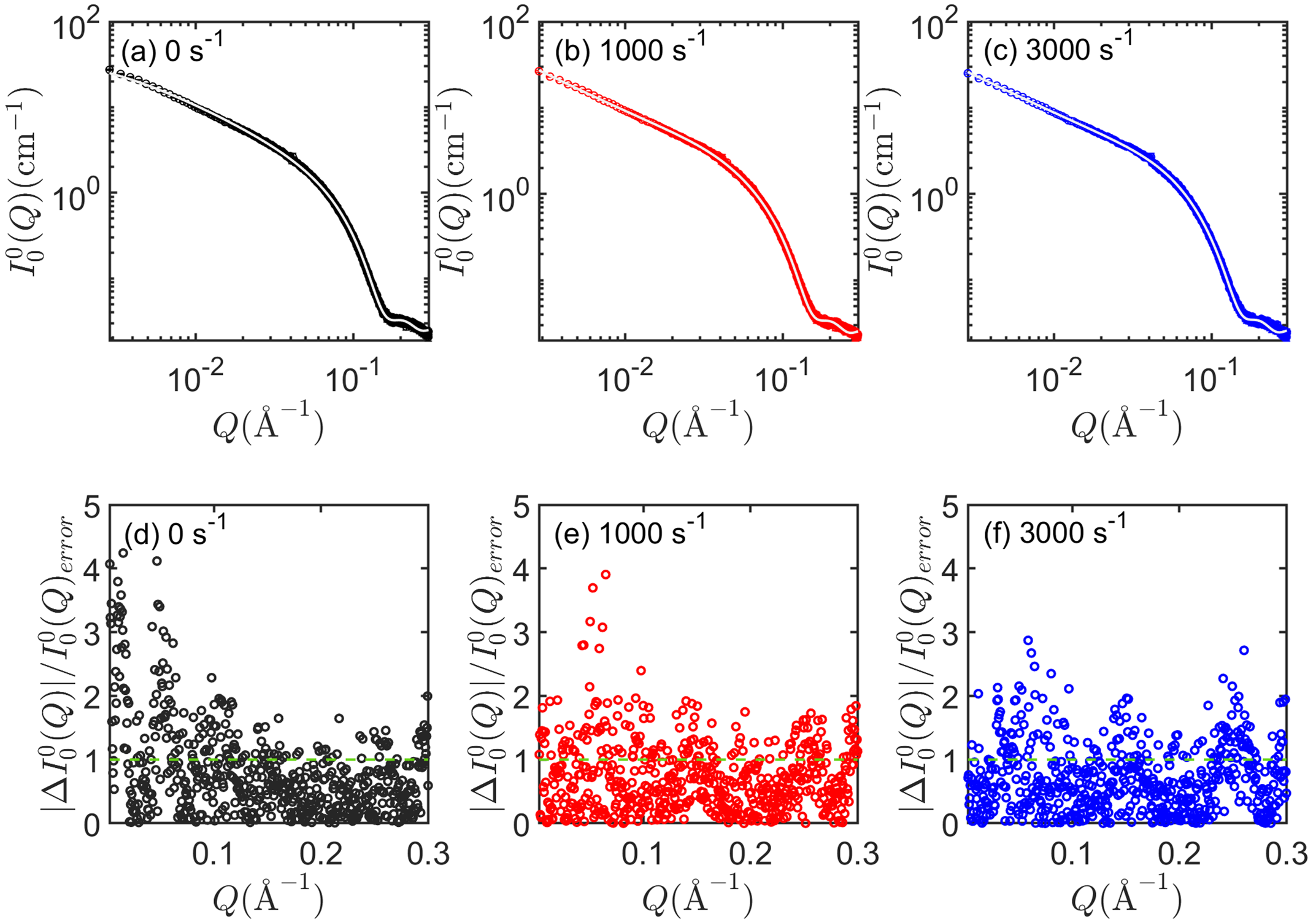}
}  
\caption{
Comparative analysis of the experimentally obtained \( I_0^0(Q) \) across different states: (a) quiescent, (b) under shear at \(\dot{\gamma} = 1000 \, \text{s}^{-1}\), and (c) under shear at \(\dot{\gamma} = 3000 \, \text{s}^{-1}\) (colored symbols). Model fits based on Eqn.~\eqref{eq:4.51} are shown with white lines, demonstrating strong agreement with the experimental data. In panels (a)-(c), the cross-sectional radius \( R \) and the polydispersity ratio \( p_R \) are found to be approximately \( 21.5 \, \text{\AA} \) and \( 14.5 \times 10^{-2} \), respectively.  
Panels (d-f) present the normalized residuals, $\frac{|\Delta I_0^0(Q)|}{I_0^0(Q)_{error}}$, showing that fluctuations primarily occur within one standard deviation, thereby confirming the fitting accuracy and supporting the model representation of micellar conformation as a collection of polydisperse rods.
}  
\label{fig:7}
\end{figure}

The form factor \(P(Q, L, R)\) for a uniform cylinder with radius \(R\) and length \(\overline{L}\) is given by the following expression:

\begin{eqnarray}
P(Q, L, R) &=& 
V_p^2 \int_{0}^{1} d\mu \left[ \frac{\sin \left( \frac{1}{2} Q L \mu \right)}{\frac{1}{2} Q L \mu} \right]^2 
\left[ \frac{2J_1(QR \sqrt{1 - \mu^2})}{QR \sqrt{1 - \mu^2}} \right]^2.
\label{eq:4.4}
\end{eqnarray}
In this expression, \( V_p = \pi R^2 L \) denotes the cylindrical volume, \( J_1 \) is the Bessel function of the first kind of order \( 1 \), and \( \mu \) is a dummy integration variable. According to the literature, the distribution of micellar lengths can be well-described by the Schulz distribution \cite{Schulz, Huang5, Cates2, Stukalin2006}:

\begin{equation}
f(L) = \frac{1}{\Gamma(z+1)} \left( \frac{z+1}{\overline{L}} \right)^{z+1} L^z \exp\left( -\frac{z+1}{\overline{L}} L \right).
\label{eq:4.5}
\end{equation}
Here, \( \overline{L} \) is the average rod length, and the parameter \( z \) controls the width of the distribution. The experimental scattering data reveal that the first minimum in SANS intensities is smeared due to the slight polydispersity in the cross-sectional radius \( R \). This effect can be accurately modeled using a central moment expansion \cite{Huang4}:

\begin{equation}
I_0^0(Q) = A \int dL\, f(L) \left[ P(Q, L, \overline{R}) + \frac{(\overline{R} p_R)^2}{2} \frac{d^2 P(Q, L, \overline{R})}{d \overline{R}^2} \right]  + I_{\text{inc}},
\label{eq:4.51}
\end{equation}
where \( p_R \) is the ratio of the standard deviation in \( \overline{R} \) to \( \overline{R} \). In Figs.~\ref{fig:7}(a-c), we present three comparative examples of the experimentally obtained \(I_0^0(Q)\) (represented by colored symbols) in different states: (a) quiescent, (b) at a shear rate of \(\dot{\gamma} = 1000 \, \text{s}^{-1}\), and (c) at \(\dot{\gamma} = 3000 \, \text{s}^{-1}\). The corresponding model fits based on Eqn.~\eqref{eq:4.51} are illustrated with white lines. The quantitative agreements between the experimental data and model predictions are visually evident.

Figs.~\ref{fig:7}(d-f) display the results of $\frac{|\Delta I_0^0(Q)|}{I_0^0(Q)_{error}}$, a commonly used statistical method for error estimation in linear regression analysis \cite{Demidenko}, where $\Delta I_0^0(Q)$ is the difference between the experimentally measured and model fitting isotropic scattering intensities and $I_0^0(Q)_{error}$ is the experimental uncertainty of isotropic scattering intensity.  Within the probed range of \(Q\), the majority of the results are observed to fluctuate around one standard deviation. This analysis further corroborates the quality of the fitting, suggesting that the overall micellar conformation can be accurately represented by a collection of polydisperse rods.

\begin{figure}[h!]
\centerline{
  \includegraphics[width=\linewidth]{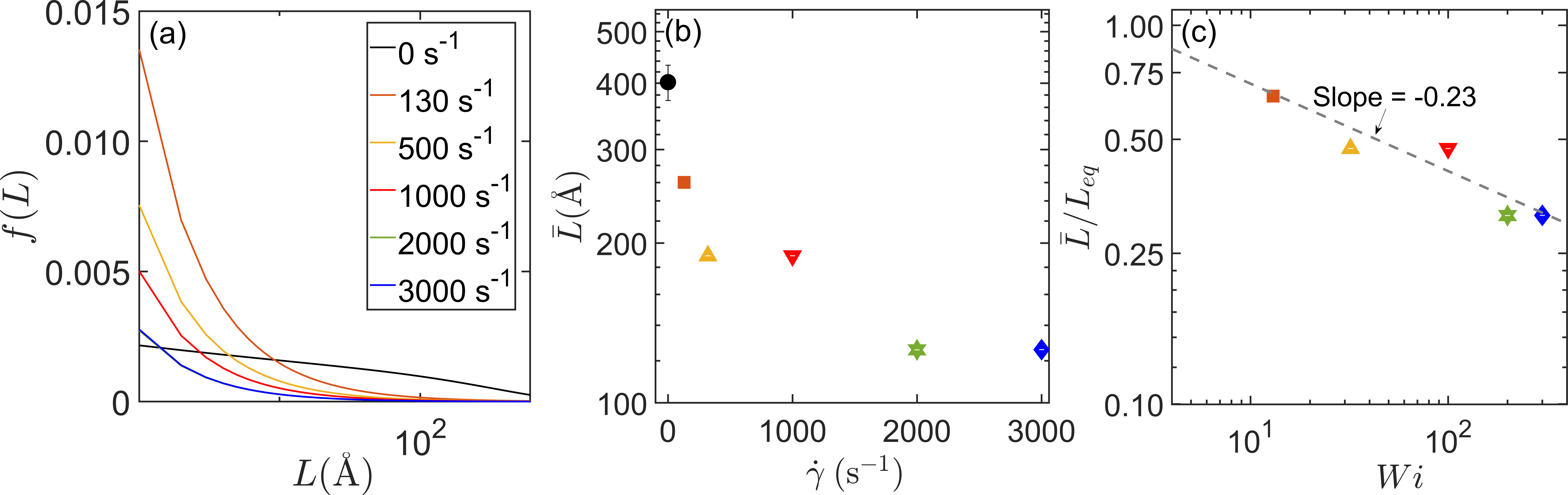}
}  
\caption{Length distribution function \( f(L) \) and average micellar length \( \overline{L} \) derived from model fitting analysis, illustrating the statistical distribution of micellar lengths. As the shear rate \( \dot{\gamma} \) increases from 0 to 3000 $\text{s}^{-1}$ (a), both the width and skewness of \( f(L) \) decrease, indicating enhanced uniformity and symmetry in micellar lengths. The average micellar length \( \overline{L} \) is calculated as \( \overline{L} = \int L f(L) dL \). \textcolor{black}{Note that the \( f(L) \) for \( \dot{\gamma} = 2000 \, \text{s}^{-1} \) essentially overlaps with that for \( \dot{\gamma} = 3000 \, \text{s}^{-1} \).} Results (b) show a significant reduction in \( \overline{L} \) from 400 $\text{Å}$ to 150 $\text{Å}$, highlighting the dynamic nature and conformational integrity of micelles under flow. \textcolor{black}{(c) Comparison of neutron scattering experiments and simulations: $\frac{\overline{L}}{L_{eq}}$, the normalized average micellar length ($L_{eq}$ is the quiescent-state value, Fig.~\ref{fig:8}(b)), is plotted against the Weissenberg number $W_i$. Micellar scission occurs for $1 < W_i < 10$, with $\frac{\overline{L}}{L_{eq}} \sim 0.7$ at $W_i = 10$, consistent with dissipative particle dynamics (DPD) simulations \cite{Koide1}. At $W_i \sim 100$, $\frac{\overline{L}}{L_{eq}} \sim \frac{1}{3}$, matching DPD results. The slope of $-0.23$ (dashed line) closely aligns with the DPD value of $-0.25$, demonstrating strong quantitative agreement.}
}  
\label{fig:8}
\end{figure}

In Fig.~\ref{fig:8}, we present the length distribution function \( f(L) \) alongside the average micellar length \( \overline{L} \), both derived from our model fitting analysis. The length distribution function \( f(L) \) characterizes the statistical distribution of micellar lengths within the sample, offering valuable insights into the variability and behavior of the micellar structures. Meanwhile, the average micellar length \( \overline{L} \), extracted from this distribution, serves as a crucial parameter for understanding micellar dynamics and their implications for the rheological properties of the system under investigation. Together, these metrics provide a comprehensive perspective on the micellar population and their responses to flow conditions.

As illustrated in Fig.~\ref{fig:8}(a), increasing the shear rate \( \dot{\gamma} \) from 0 to 3000 $\text{s}^{-1}$ leads to a consistent decrease in both the width and skewness of \( f(L) \). This reduction in width signifies decreased variability and a more concentrated size distribution around the mean, indicating enhanced uniformity in micellar lengths. Concurrently, the decline in skewness reflects diminished asymmetry, with values approaching zero suggesting that the distribution is becoming increasingly symmetric. As the skewness approaches zero, the Schulz distribution trends toward a Gaussian distribution, indicating a balanced presence of rods with both shorter and longer lengths. Overall, the decline in both width and skewness underscores the growing uniformity and symmetry of the particle size distribution, emphasizing the enhanced consistency of micellar lengths.

Given the length distribution \( f(L) \) in Fig.~\ref{fig:8}(a), the average micellar length \( \overline{L} \) can be computed using the following expression:

\begin{equation}
\overline{L} = \int L f(L) dL  
\label{eq:4.6}
\end{equation}
The results displayed in Fig.~\ref{fig:8}(b) quantitatively characterize the flow-induced scission in rod-like micelles, revealing a significant reduction in \( \overline{L} \) from 400 $\text{Å}$ in the quiescent state to 150 $\text{Å}$. This dramatic decrease not only underscores the kinetic nature of micellar structures under flow but also highlights the critical impact of shear on their conformational integrity and behavior.

\textcolor{black}{It is instructive to compare the experimental observation of flow-induced micellar scission with findings from computational studies.} 

\textcolor{black}{To compare neutron scattering experiments with simulations, $\frac{\overline{L}}{L_{eq}}$, where $L_{eq}$ is the average micellar length in the quiescent state (indicated by the black symbol in Fig.~\ref{fig:8}(b)), is plotted against the Weissenberg number $W_i$ in Fig.~\ref{fig:8}(c). In line with dissipative particle dynamics (DPD) simulations of unentangled surfactant micelles by Koide and Goto \cite{Koide1}, micellar scission occurs in the range $1 < W_i < 10$, with $\frac{\overline{L}}{L_{eq}} \sim 0.7$ at $W_i = 10$, and simulations predicting scission for $W_i \geq 4$. At $W_i \sim 100$, $\frac{\overline{L}}{L_{eq}}$ decreases to approximately $\frac{1}{3}$, agreeing with DPD results. The decline in $\frac{\overline{L}}{L_{eq}}$ with increasing $W_i$ follows a slope of about $-0.23$ (dashed line), closely matching the DPD simulation slope of $-0.25$. These results underscore the strong quantitative agreement between our SANS measurements and DPD simulations.} 

\textcolor{black}{These findings highlight the agreement between experiments and simulations, bridging theoretical predictions and real-world behavior. The results in Fig.~\ref{fig:8} demonstrate that neutron scattering experiments provide quantitative validation of flow-induced micellar scission. By enabling precise comparisons with computational studies, this work advances our understanding of micellar dynamics and underscores the critical role of experiments in refining theoretical models. The synergy between experiment and simulation reinforces the robustness of this approach and opens avenues for exploring complex flow-driven processes in soft matter systems.}

\textcolor{black}{In addition to providing the first quantitative experimental insights into micellar scission dynamics, this scattering study raises important questions about the universal scalability of scission dynamics. While the DPD simulations investigated the self-assembly of nonionic surfactants, this work focuses on ionic CTAB systems. Remarkably, the experimental results shown in Fig.~\ref{fig:8}(c) quantitatively align with the DPD simulations. This agreement underscores the broader applicability of the simulations and invites further investigation to determine whether this alignment reflects an authentic universal principle or is merely a coincidence.}

\subsection{\textcolor{black}{Characterizing the Orientational Distribution of Flowing Rod-like Micelles}}

\textcolor{black}{The advantage of spherical harmonic expansion in spectral analysis lies in its ability to quantify the micellar orientational distribution function by capturing the angular distribution of orientations. This method effectively separates different structural and flow-related effects in a systematic, compact, and physically interpretable manner.}

\begin{figure}[h!]
\centerline{
  \includegraphics[width=\linewidth]{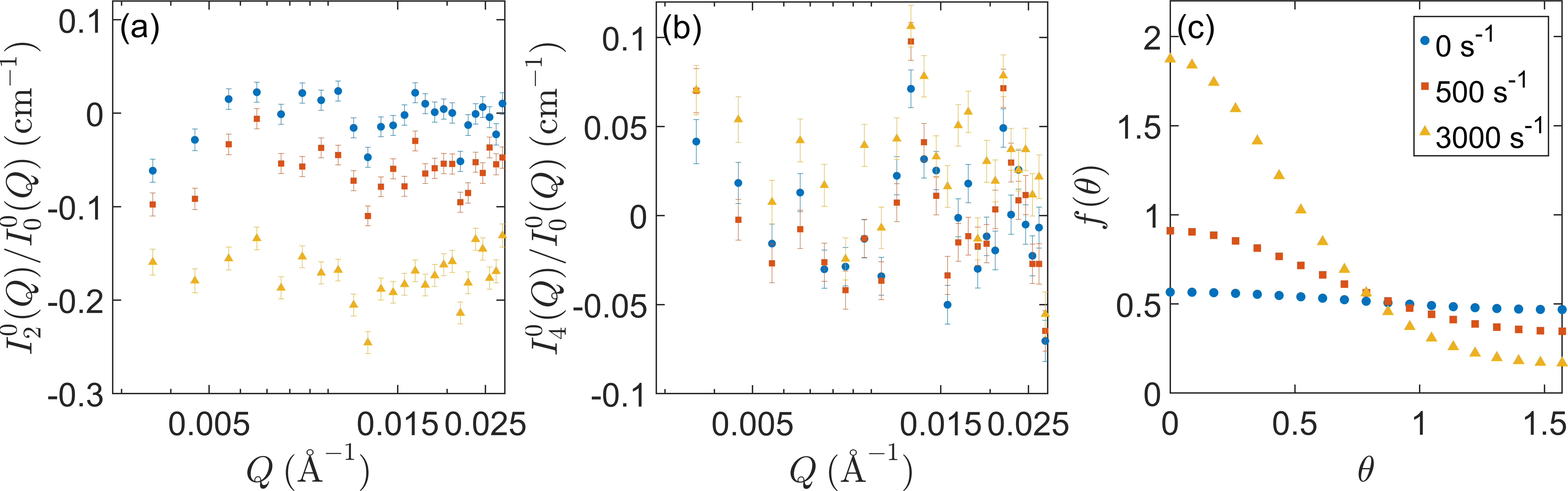}
}  
\caption{The ratios: (a) \( \frac{I_2^0(Q)}{I_0^0(Q)} \) and (b) \( \frac{I_4^0(Q)}{I_0^0(Q)} \) \textcolor{black}{at different} shear rates are obtained using the real spherical harmonics decomposition of the experimentally measured scattering intensity \( I(\mathbf{Q}) \). Obviously, \( \frac{I_4^0(Q)}{I_0^0(Q)} \) is approximately an order of magnitude smaller than \( \frac{I_2^0(Q)}{I_0^0(Q)} \), indicating that \( I_2^0(Q) \) dominates the scattering anisotropy of \( I(\mathbf{Q}) \).  Using the principle of maximum entropy, Eqns. (12) and (13), with the approximated orientation order parameter \(S_2^0\) as the average value of \( -\frac{2I_2^0(Q)}{I_0^0(Q)} \), the reconstructed orientation distribution function \( f(\theta)\) is displayed in panel (c).}   
\label{fig:9}
\end{figure}

The orientational information is encoded in the anisotropic components of the RSH \cite{Huang2, Kardar}. To illustrate this, we performed an eigen-decomposition of the spectra using Eqn.~\eqref{eq:3.1} and extracted the anisotropic components, namely $I_2^0(Q)$ and $I_4^0(Q)$, as shown in Fig.~\ref{fig:9}. Deviations from the equilibrium configuration, characterized by random orientation, were assessed by normalizing the results with the magnitude of the isotropic component, $I_0^0(Q)$.

As the shear rate increases, the ratio \( I_2^0(Q)/I_0^0(Q) \) decreases distinctly from the asymptotic isotropic value of 0 in the quiescent state to \(-0.2\) at a shear rate of 3000 \,s\(^{-1}\), as shown in Fig.~\ref{fig:9}(a). In contrast, the ratio \( I_4^0(Q)/I_0^0(Q) \) remains close to zero, exhibiting minimal dependence on the shear rate, as depicted in Fig.~\ref{fig:9}(b). These results emphasize that, within the context of this experiment, \( I_2^0(Q) \) is the dominant anisotropic component for characterizing the orientational distribution function (ODF), \( f(\theta) \), of sheared rod-like micelles.

To reconstruct $f(\theta)$, we applied the principle of maximum entropy \cite{Huang3} with $I_2^0(Q)$ as input. The ODF, $f(\theta)$, is expressed as
\begin{equation}
    f(\theta) = A_0 \exp[A_2^0 Y_2^0(\theta, \phi)],
\end{equation}
{\color{black} where $A_0$ and $A_2^0$ are constants to be determined. Since the ratios $I_2^0(Q)/I_0^0(Q)$ approximate a constant and the system under investigation consists of rod-like micelles, the value of $-2I_2^0(Q)/I_0^0(Q)$ corresponds to the orientational order parameter $S_2^0$ \cite{Huang3}.}

Additionally, since \( f(\theta) \) is normalized, it can be calculated using the following \textcolor{black}{integrals}:
\begin{equation}
    \int_0^\pi d\theta \sin\theta f(\theta) = 1, \quad \int_0^\pi d\theta \sin\theta f(\theta) Y_2^0(\theta,\phi) = S_2^0.
\end{equation}

The reconstructed results of \( f(\theta) \) at different shear rates are presented in Fig.~\ref{fig:9}(c), where \( S_2^0 \) is taken as the average value of \( -2 I_2^0(Q)/I_0^0(Q) \) within the probed \( Q \) range. It is instructive to point out that the rod lengths in the system are highly polydisperse. In the intermediate \( Q \) regime, the rod orientations are coupled with their lengths, which is reflected in the \( Q \)-dependent ratios of \( I_2^0(Q)/I_0^0(Q) \).

\textcolor{black}{Since the 80's \cite{Hayter3, Hayter1, Hayter2}, many scattering studies on cylindrically elongated micellar systems under flow have been published. Most rheo-SANS investigations primarily focused on the flow-induced alignment of these systems. Quantitative analysis in these studies typically involves two main approaches: parametric regression analysis, which assumes \textit{a priori} confidence in the analytical expression of the orientation distribution based on pre-selected parameters, with the numerical values determined via regression techniques \cite{Hayter3, Kalus1, Kalus2, Kalus3, Hayter1, Hayter2, Foerster, Tabor2, Helgeson}, and the scalar descriptor method, which uses simpler metrics, such as the order parameter or empirical alignment factor, to quantify alignment \cite{Wagner1, Wagner2, Wagner3, Wagner4, Yun, Weigandt, Lionel2, Helgeson}. While both approaches provide valuable insights into flow-induced alignment, they both fail to capture micellar scission, either due to the complexity of modeling the mechanically driven micellar conformation or the inherent limitations of these methods. The key challenge is their inability to simultaneously account for the structural changes in elongated micellar systems, such as the evolution of orientation, scission, and changes in size distribution.}

\textcolor{black}{The spherical harmonic decomposition method proposed in this work aims to overcome these limitations by offering a more nuanced structural representation of micelles under mechanical stress. This method enables the simultaneous consideration of alignment and structural changes, including scission, under flow. By disentangling these factors, we gain a more comprehensive understanding of the system’s behavior. Our approach not only quantifies flow-induced fragmentation and length reduction but also provides detailed insights into the evolving length distribution and orientation of micelles. To our knowledge, this is the first comprehensive and robust method to disentangle hydrodynamic and structural effects associated with shear deformation. Early studies correlating the non-monotonic dependence of the radius of gyration ($R_G$) with micellar size were insufficient for fully capturing the variations in both size and orientation \cite{Warr2}. A more sophisticated data analysis approach, utilizing the gyration tensor \cite{Huang1}, particularly its second moment of density fluctuations along different spatial directions, offers a more complete characterization of the system's anisotropic properties, with $R_G$ serving as a trace of the gyration tensor.}

\textcolor{black}{Our advancements in data analysis, coupled with optimized shear cell geometry, establish a new benchmark for scattering-based analyses of complex micellar systems, providing a robust framework for understanding the interplay between flow dynamics and micellar behavior.}

\section{Conclusions}
\label{sec:5}

In summary, this study \textcolor{black}{offers a comprehensive investigation into the structural response of rod-like micelles under shear flow, uncovering critical insights into flow-induced scission mechanisms and micellar conformational behavior \textcolor{black}{ that only computer simulation have so far eluded}. By applying the spherical harmonic decomposition approach, we have performed a quantitative and unbiased analysis of scattering data, enhancing the detection of subtle structural transformations. Key findings from our spectral analysis include the steady decrease in isotropic scattering intensity, \(I_0^0(Q)\), and the reduction in average micellar length from 400 \AA{} in the quiescent state to 150 \AA{} under high shear \textcolor{black}{in complete agreement with recent DPD simulation predicting micelle shortening when $1 < W_i < 100$}. These results highlight the dynamic nature of rod-like micelles in shear flow, with direct implications for their colloidal stability and industrial applications.}

\textcolor{black}{Our hypothesis is based on the premise that micellar scission under mechanical stress can be accurately captured by isolating the isotropic component \(I_0^0(Q)\), providing a quantitative framework for evaluating flow-induced structural changes. The innovation in our approach lies in applying spherical harmonic decomposition within the flow-velocity gradient plane, enabling precise tracking of changes in micellar length distribution. Furthermore, the anisotropic component reveals enhanced alignment of the flow-induced shortened rod-like micelles. This methodology marks a significant improvement over previous scattering studies, which often lacked the resolution to identify these specific transformations under high shear. Compared to existing literature relying on approximate or indirect measurements of micellar deformation \cite{Hayter1, Wagner3, Rothstein2, Tabor2, Helgeson}, our model-free spectral decomposition approach allows direct quantification of length and orientational distribution, establishing a robust framework for understanding micellar responses to shear.}

\textcolor{black}{Looking ahead, this method could be extended to explore a broader range of soft matter systems beyond rigid rod-like micelles, such as flexible cylindrically elongated micelles or fibrillar aggregates, where shear flow and mechanical forces drive complex morphological transformations. In studying the influence of the interplay between thermodynamic and fluid mechanical interactions on the conformation of elongated micelles with flexible configurations, conformational characteristics such as contour length and bending energy could be directly calculated from the extracted \(I_0^0(Q)\). Since the orientational distribution for such flexible micelles would exhibit a dependence on length scale, one could envision extending the approach developed in this work to generalize the alignment function $f(\theta)$, as displayed in Fig.~\ref{fig:9}(c), to a more comprehensive form $f(\theta,L)$, suitable for describing the alignment of flexible, elongated objects. These advancements hold promising potential to deepen our understanding of complex soft matter systems and to reveal new mechanisms underlying their dynamic behavior under external forces.}

Future work should aim to refine the spherical harmonic decomposition model to accommodate more intricate micellar structures and develop predictive models for their behavior under varying flow conditions. This approach lays a foundational framework for these studies, with potential applications in optimizing formulations for catalysis, nanomedicine, and soft materials engineering.

\clearpage
\section{Acknowledgement}
\label{sec:6}

A portion of this research used resources at the Spallation Neutron Source, a DOE Office of Science User Facility operated by the Oak Ridge National Laboratory. This research was sponsored by the Laboratory Directed Research and Development Program of Oak Ridge National Laboratory, managed by UT-Battelle, LLC, for the U. S. Department of Energy. RPM acknowledges support from CHRNS, a national user facility jointly funded by the NCNR and the NSF under Agreement No. DMR-2010792. Commercial equipment or software identified in this work does not imply recommendation nor endorsement by NIST. GRH was supported by the National Science and Technology Council in Taiwan with Grant No. NSTC 111-2112-M-110-021-MY3. We would like to thank ILL for the provision of beam time on the D22 SANS instrument  DOI: 10.5291/ILL-DATA.EASY-1399.  

\clearpage
 \bibliographystyle{elsarticle-num} 

\end{document}